\documentclass[referee,sn-basic]{sn-jnl}

\usepackage{graphicx}%
\usepackage{multirow}%
\usepackage{amsmath,amssymb,amsfonts}%
\usepackage{amsthm}%
\usepackage{mathrsfs}%
\usepackage[title]{appendix}%
\usepackage{xcolor}%
\usepackage{textcomp}%
\usepackage{manyfoot}%
\usepackage{booktabs}%
\usepackage{algorithm}%
\usepackage{algorithmicx}%
\usepackage{algpseudocode}%
\usepackage{listings}%

\usepackage{hyperref}
\usepackage{orcidlink}
 \usepackage{lineno}

\usepackage{MnSymbol}
	
\def\bmath{}

\def\centerline{}

\usepackage{natbib}
\setcitestyle{authoryear,open={(},close={)}}

\theoremstyle{thmstyleone}%
\newtheorem{theorem}{Theorem}
\newtheorem{lemma}[theorem]{Lemma}
\theoremstyle{thmstyletwo}%
\newtheorem{example}{Example}%

\theoremstyle{thmstylethree}%
\newtheorem{definition}{Definition}%

\begin{document}

\title[Multiple testing of interval composite null hypotheses using randomized $p$-values]{Multiple testing of interval composite null hypotheses using randomized $p$-values}


\author[]{\fnm{Daniel} \sur{Ochieng\orcidlink{0000-0002-1023-5028}}}\email{dochieng@uni-bremen.de}



\affil[]{\orgdiv{Institute for Statistics}, \orgname{University of Bremen}, \orgaddress{
\postcode{28344}, \state{Bremen}, \country{ Germany}}}




\abstract{One class of statistical hypothesis testing procedures is the indisputable equivalence tests, whose main objective is to establish practical equivalence rather than the usual statistical significant difference. These hypothesis tests are prone in ``bioequivalence studies," where one would wish to show that, for example, an existing drug and a new one under development have the same therapeutic effect. In this article, we consider a two-stage randomized (RAND2) $p$-value utilizing the uniformly most powerful (UMP) $p$-value in the first stage when multiple two-one-sided hypotheses are of interest. We investigate the behavior of the distribution functions of the two $p$-values when there are changes in the boundaries of the null or alternative hypothesis or when the chosen parameters are too close to these boundaries. We also consider the behavior of the power functions to an increase in sample size. Specifically, we investigate the level of conservativity to the sample sizes to see if we control the $\alpha$ level when using either of the two $p$-values for any sample size. In multiple tests, we evaluate the performance of the two $p$-values in estimating the proportion of true null hypotheses. We conduct a family-wise error rate control using an adaptive Bonferroni procedure with a plug-in estimator to account for the multiplicity that arises from the multiple hypotheses under consideration. We verify the various claims in this research using simulation study and real-world data analysis.
}






\maketitle

\section{Introduction}

Equivalence tests are testing procedures for establishing practical equivalence rather than the usual statistical significant difference. Within the frequentist framework, this test uses the fact that failing to reject a given null hypothesis of no difference is not reasonably equivalent to accepting the null hypothesis. Equivalence studies are common in the medical field, for example, where one would wish to show that an existing drug and a new one under development have the same therapeutic effect. We refer to such studies as ``bioequivalence studies" and classify them into three categories according to the distance measure between two populations. The categories are individual, population, and average equivalence. Another common area of application is in genetics, where they can be used to identify non-DE (differentially expressed) genes (cf. \cite{qiu2010evaluation}) or to test for Hardy-Weinberg equilibrium (HWE) in the case of multiple alleles as in \cite{ostrovski2020new}. Other areas of application of equivalence tests include the comparison of similarity between two Kaplan-Meier curves, which estimate the survival functions in two populations. See Sect. 1.3 of \cite{wellek2010testing} for an in-depth discussion of these applications. We can state equivalence as a difference or ratio between two means. Rejecting the null hypothesis is the same as declaring an equivalence. This rejection is similar to the interval under the alternative hypothesis containing a zero (for the difference in means hypothesis) or a one (for the ratio of means hypothesis).

Some studies on equivalence testing include \cite{romano2005optimal}, which provides bounds for the asymptotic power of equivalence tests and constructs efficient tests that attain those bounds. The same author also gives an asymptotically UMP test based on Le Cam's notion of convergence of experiments for testing the mean of a multivariate normal. Equivalence tests can also use intersection-union tests (cf. \cite{berger1996bioequivalence}) since the null is a union of several null hypotheses, and the alternative is an intersection of many rejection regions. \cite{berger1996bioequivalence}  consider this intersection-union test for the simultaneous assessment of equivalence on multiple endpoints. This test requires that all the $(1-2\alpha)100\%$ simultaneous 
intervals fall within the equivalence bounds for an overall $\alpha$ level test. This approach can be conservative depending on the correlation structure among the endpoints and the study power. Another popular approach to equivalence testing is the Two One-Sided Test (TOST) procedure. We can use this procedure as an alternative to the goodness-of-fit tests. However, since it is sensitive to the noise level in the data, it can have low power for data sets with a high variance. Alternatively, we can use a distance measure such as the Euclidean distance between two probability vectors.

\cite{munk1996equivalence} considered equivalence tests for Lehmann's alternative, which are unbiased for equal sample sizes within the two groups. An extension of the expected $p$-value of a test (EPV) to univariate equivalence tests was considered by
\cite{pfluger2002assessing}. Since this procedure is independent of the distribution of the test statistic under the null hypothesis, it avoids the problem of looking for this distribution for the test statistic. Furthermore, the EPV is independent of the nominal level $\alpha.$

Equivalence tests are univariate, and we can apply them to each characteristic of interest without a multiplicity adjustment. 
However, the probability of making false claims of equivalence (type I errors) increases when we analyze multiple characteristics without a multiplicity adjustment. \cite{leday2023improved} proposed a familywise error rate (FWER) control based on Hochberg’s method. The same authors also showed that Hommel’s method performs as well as Hochberg’s and that an “adaptive” version of Bonferroni’s method outperforms Hommel’s in-terms of power for equivalence testing. \cite{giani1991some} and \cite{giani1994testing} on the other hand considered simultaneous equivalence tests in the  $k-$sample case  and proposed tests based on the  range statistic.  \cite{qiu2010evaluation} and \cite{qiu2014applying} consider multiple equivalence tests based on the average equivalence criterion to identify non-DE genes. Both articles investigate the power and false discovery rate (FDR) of the TOST. Since the variance estimator in the TOST procedure can become unstable and lead to low power for small sample sizes, the later article proposes a shrinkage variance estimator to improve the power. \cite{huang2006statistical} also applied an average equivalence test criterion but adjusted for the multiplicity using the simultaneous confidence interval approach.


Multiple test procedures that utilize $p$-values are valid only if the $p$-value statistics are uniformly $(0,1)$ distributed under the null hypothesis. Since we use the $p$-values many times,  any non-uniformity in their distribution quickly accumulates and reduces the power of the overall procedure. We can decompose the equivalence hypothesis into two one-sided hypotheses, each leading to a composite null hypothesis. The $p$-values from such a hypothesis can fail to follow the uniform distribution if we do not compute them under the least favorable parameter configurations (LFCs). Furthermore, we can have categorical data, for example, in genetic association studies that generate discrete data in counts, leading to test statistics with discrete distributions. Since the $p$-value is a deterministic transformation of the test statistic, this leads to discretely distributed $p$-values that are also nonuniform under the null hypothesis.

The problems of composite nulls and discrete test statistics can lead to conservative $p$-values, which implies that the $p$-value is stochastically larger than $UNI(0,1)$ distribution under the null hypothesis. To our knowledge, no research has previously considered a two-stage randomized $p$-values in testing for equivalence hypotheses. In this article, we propose a two-stage randomized $p$-value for multiple testing of equivalence hypotheses to address these two issues. The two-stage procedure uses the UMP $p$-value in the first stage to remove the discreteness of the test statistic. The randomized $p$-value proposed in \cite{hoang2021usage} for a continuous test statistic is then used in the second stage to deal with the composite null hypothesis.


When utilizing the non-randomized version of the Two One-Sided Test (TOST) UMP $p$-value in discrete models, \cite{finner2001increasing} showed that it is possible for the power function based on a sample of size $n$ to coincide on the entire parameter space with the corresponding power function based on size $n+i$ for small $i\in \mathbb{N}$. We illustrate that the power function of a test based on the two-stage randomized (RAND2) $p$-value for discrete models, just like the one for the UMP randomized $p$-value, is strictly increasing with an increase in the sample size. We further illustrate that for small sample sizes, it is possible that the power functions of the test based on the two $p$-values (UMP and RAND2) do not strictly increase with an increase in the sample size. 

We also investigate the behavior of the distribution function for the UMP and RAND2 $p$-values under the null and alternative hypothesis. Three objectives are of interest: First, to find if the power and level of conservativity of the $p$-values depend on the size of the equivalence limit. Second, to investigate the behavior of the CDFs when the chosen parameter is close to the boundary of the null or alternative hypothesis, and third, to find out if the level of conservativity of the $p$-values depends on the sample sizes. Finally, we consider multiple testing of equivalence hypotheses where we assess the performance of our $p$-values in estimating the proportion of true null hypotheses using an empirical-CDF-based estimator. An adaptive version of the Bonferroni that utilizes the plug-in estimator of \cite{finner2009controlling} is used for familywise error control.


The rest of this paper is organized as follows. General preliminaries are provided in Section $\ref{general-assum}$. The definitions, CDFs, and investigations of the behaviors of those CDFs under the null and alternative hypothesis for the UMP and the two-stage randomized $p$-values are considered in Section $\ref{interval_composite}$. We also investigate if the power function of the $p$-values is monotonically increasing with an increase in the sample size in the same section. Furthermore, we give the parameter value that maximizes the power of a test based on the $p$-values in the same section. We defer all matters concerning multiple testing until Section \ref{multiple_interval}, where we consider a real-world data analysis and a simulation study to assess the performance of the $p$-values in estimating the proportion of true null hypotheses. Finally, we discuss our results and give recommendations for future research in Section $\ref{s:discuss}$.

\section{General preliminaries}\label{general-assum}

Let $\pmb{X}=(X_1,\ldots,X_n)^\top$  denote our random data where each $X_r$ is a real-valued, observable random variable, $1 \leq r \leq n$ with the support of $\pmb{X}$  denoted by $\mathcal{X}$. We assume all $X_r$ are stochastically independent and identically distributed (i.i.d.) with a known parametric distribution. The marginal distribution of $X_1$ is assumed to be $P_\theta$, where $\theta \in\Theta \subseteq \mathbb{R}$ is the model parameter. The distribution of $\pmb{X}$ under $\theta$ is as a result given by $P_\theta^{\otimes n} =: \mathbb{P}_\theta$. We will be concerned with an interval hypothesis test problem of the form
\begin{equation}
H: \theta \notin (\theta_1, \, \theta_2)\ \text{~~versus~~} \  K: \theta \in (\theta_1, \, \theta_2),
\label{testproblem}
\end{equation} 
for given numbers $\theta_1,\theta_2 \in \Theta$ such that $\theta_1<\theta_2$. When $k$ hypotheses are of interest, then they will be expressed as $H_j: \theta\notin \Delta_j$ versus $K_j:\theta\in \Delta_j$ where $\Delta_j$ denotes the range of values in the $j^{th}$ interval between $\theta_1^{(j)}$ and $\theta_2^{(j)}$ for $j\in \{1,\ldots,k\}$ and $k$ is the multiplicity of the problem. Denote the resulting $k$ $p$-values by $p_1,\ldots,p_k$. We consider the case $k=1$ in Section \ref{interval_composite} and defer the multiple test problem till Section \ref{multiple_interval}.  When the difference between the $j^{th}$ true parameter $\theta^{(j)}$ and $\theta_1^{(j)}$ or $\theta_2^{(j)}$ ($j=1,\ldots,k$) is kept constant for all the $k$ hypotheses, then this is referred to as the ``average equivalence" criterion. We can sometimes make the interval in \eqref{testproblem} symmetric to achieve equivariance to the permutation of groups, for example, the choice $\theta_2=\theta_1^{-1}$ in \cite{pfluger2002assessing} and \cite{munk1996equivalence}. 

As mentioned before, one method for testing this hypothesis is the Two One-Sided Test (TOST) procedure, where one tests for the alternatives $\theta<\theta_1$ and $\theta>\theta_2$ separately at size $\alpha$ and in no particular order. TOST is a particular case of the intersection-union test proposed by \cite{berger1982multiparameter}  where the null hypothesis is a union of disjoint sets, and the alternative hypothesis is an intersection of the complements of those sets. For this reason, we conduct the separate individual tests at size $\alpha$ without a multiplicity adjustment like $\alpha/2$. Practical equivalence is declared if one rejects both tests and otherwise non-equivalence. These procedures suffer from a lack of power, and an alternative that is more powerful but too complicated has been suggested in the literature by \cite{berger1996bioequivalence} and \cite{brown1997unbiased}. Since alternative tests are difficult to implement, we use TOST in this research.

We consider test statistics $T(\pmb{X})$, where $T: \mathcal{X} \to \mathbb{R}$ is a measurable mapping. Furthermore, the test statistics $T_r$ for $r=1,\ldots,n$ are also assumed to be mutually independent. 
The marginal $p$-value $p(\pmb{X})$ resulting from $T(\pmb{X})$ 
is assumed to be valid, meaning that $\mathbb{P}_\theta(p(\pmb{X}) \leq \alpha)\leq \alpha$ holds true for all $\alpha \in[0,1]$ and for any parameter value $\theta$ in the  null hypothesis.
Valid $p$-values are stochastically larger than UNI $(0,1)$, as investigated by, among many others, \cite{habiger2011randomised} and \cite{dickhaus2012analyze}. On the same note, we call a $p$-value conservative if it is valid and  $\mathbb{P}_\theta(p(\pmb{X}) \leq \alpha)<\alpha$ holds true for some $\alpha \in(0,1)$. Throughout the article, we refer to the CDF of a $p$-value under the alternative hypothesis as a power function because we reject the null hypothesis for small $p$-values. Finally, we also make use of the (generalized) inverses of certain non-decreasing functions mapping from $\mathbb{R}$ to $[0, 1]$. In this regard, we follow Appendix 1 in \cite{reiss1989}: If $F$ is a real-valued, non-decreasing, right-continuous function, and similarly $G$ is a real-valued, non-decreasing, left-continuous function where we define both $F$ and $G$ on $\mathbb{R}$, then  $F^{-1}(y) = \inf\{x \in \mathbb{R}: F(x) \geq y\}$ and $G^{-1}(y) = \sup\{x \in \mathbb{R}: G(x) \leq y\}$, respectively.

\section{Interval composite hypothesis}
\label{interval_composite}

\subsection{Introduction}\label{interval_composite_intro}

In this article, we are interested in the (interval) composite null hypothesis of the form in \eqref{testproblem}. We test this hypothesis using two different $p$-values whose definitions and the CDFs we now give as follows.  
\begin{definition}[First stage randomization]
\label{def_ump_p_value}
Let $U$ be a UNI$ (0,1)$-distributed random variable independent of the data $\pmb{X}$. Further assume that $T(\pmb{X})$ is our test statistic whose distribution has monotone likelihood ratio (MLR), the UMP-based $p$-value $P^{UMP}(\pmb{X}, U)$ is
\begin{equation}
P^{UMP}(\pmb{X}, U)=\mathbb{P}_{\theta_i}(C_n<T(\pmb{X})<D_n)+U\mathbb{P}_{\theta_i}(T(\pmb{X})=C_n)+U\mathbb{P}_{\theta_i}(T(\pmb{X})=D_n),
\label{eq:ump_p_value}
\end{equation}
for $i=1,2$ where $\theta_1, \theta_2$ such that $\theta_1<\theta_2$ are the LFC parameters and $C_n$, $D_n \in \mathbb{R}$ such that $C_n\leq D_n$ are the critical constants. The CDF of $P^{UMP}(\pmb{X},U)$ is 
\begin{equation}
   \mathbb{P}_{\theta}\{ P^{UMP}(\pmb{X})\leq t\}=\mathbb{P}_{\theta}(C_n<T(\pmb{X})<D_n)+\gamma_n\mathbb{P}_{\theta}(T(\pmb{X})=C_n)+\delta_n\mathbb{P}_{\theta}(T(\pmb{X})=D_n),
   \label{eq:ump_p_value_cdf}
\end{equation}
where $\theta$ is the chosen true parameter while $\gamma_n$ and $\delta_n$ are the randomization constants. The critical constants $C_n, D_n\in \mathbb{R}$ and the randomization constants $\gamma_n, \delta_n\in [0,1]$ are found by solving the equation $E_{\theta_i}[T(\pmb{X})]=
\alpha$ for $i=1,2$ where $T(\pmb{X})=\sum_{r=1}^nT(X_r).$ For large sample sizes, the critical and the randomization constants are $C_n=F^{-1}_{\theta_1}(1-t)$, $D_n=F^{-1}_{\theta_2}(t)$, \[\gamma_n=\dfrac{\mathbb{P}_{\theta_1}(T(\pmb{X})\leq C_n)-(1-c)}{\mathbb{P}_{\theta_1}(T(\pmb{X})=C_n)},\ \text{and}\ \delta_n=\dfrac{c-\mathbb{P}_{\theta_2}(T(\pmb{X})\leq D_n-1)}{\mathbb{P}_{\theta_2}(T(\pmb{X})=D_n)}. \]
\end{definition}


We can use the $p$-value defined in Equation \eqref{eq:ump_p_value} with models possessing monotone likelihood ratio (MLR), for example, any one-dimensional exponential family and the location family of folded normal distribution. For continuous models, the critical constants $C_n$ and $D_n$ are slightly modified, for example, by introducing the variance in the case of a normal distribution. Moreover, the randomization constants in \eqref{eq:ump_p_value_cdf} are such that $\gamma_n=\delta_n=0$ for such continuous models. Next, we give a  lemma whose proof is in the Appendix to show that the UMP $p$-value in Definition \eqref{def_ump_p_value} is the maximum of the $p$-values for a lower- and an upper-tailed test.

\begin{lemma}
For a fixed but arbitrary significance level $\alpha\in[0,1]$ and a chosen true parameter under the null hypothesis $\theta_0=\theta_1$ or $\theta_0=\theta_2$, the UMP $p$-value in Equation \eqref{eq:ump_p_value} is the maximum of the $p$-values for a lower- and an upper-tailed test.
\label{max_p_value}
\end{lemma}

In calculating the UMP $p$-value in \eqref{eq:ump_p_value}, using either  $\theta_1$ or $\theta_2$ leads to the same result for the $p$-value. The UMP $p$-value is used in the first stage of randomization to deal with the discreteness of the test statistics. We conduct a second randomization to deal with the composite null hypothesis. The second stage randomized $p$-value (RAND2) (cf. \cite{hoang2021usage}) is defined as follows.

\begin{definition}[Second stage randomization]
Let $U$ and $\tilde{U}$ be two different UNI$ (0,1)$-distributed random variables  both stochastically independent of the data $\pmb{X}$ and are also independent of each other. Assume also that we have a constant $c\in (0,1]$. The two-stage randomized $p$-value $P^{rand2}(\pmb{X},U,\tilde{U},c)$ is 
\begin{equation}
P^{rand2}(\pmb{X},U,\tilde{U},c)=\tilde{U} \pmb{1}\{P^{UMP}(\pmb{X},U)\geq c\}+P^{UMP}(\pmb{X},U)(c)^{-1}\pmb{1} \{P^{UMP}(\pmb{X},U)<c\}. 
\label{eq:rand2_p_value}
\end{equation}
where $P^{UMP}(\bmath{X},U)$ is the UMP $p$-value in the first stage as defined in Equation $\eqref{eq:ump_p_value}$. Furthermore, we define $P^{rand2}(\pmb{X},U,\tilde{U},0)=\tilde{U}$ and  $P^{rand2}(\pmb{X},U,\tilde{U},1)=P^{UMP}(\pmb{X})$. The CDF of $P^{rand2}(\pmb{X},U,\tilde{U},c)$ is
\begin{equation}
\mathbb{P}_{\theta}\{P^{rand2}(\pmb{X},U,\tilde{U},c)\leq t\}=t\mathbb{P}_{\theta}\{P^{UMP}(\pmb{X}, U)>c\} 
+\mathbb{P}_{\theta}\{P^{UMP}(\pmb{X}, U)\leq tc\}.
\label{eq:rand2_p_value_cdf}
\end{equation} 
\end{definition}

With our $p$-values so defined, we are now ready to use them to test our hypothesis. We first describe an example of a discrete model that we use to illustrate our randomized $p$-values in practice.

\begin{example}[Binomial distribution]\label{binomial_model}
Assume that our (random) data is given by $\pmb{X}=(X_1,\ldots,X_n)^\top$, where each $X_r$ is a real-valued, observable random variable, $1 \leq r \leq n$, and all $X_r$ are stochastically independent and identically distributed (i.i.d.) Bernoulli variables with parameter $\theta_i\in (0,1)$ for $i=1,2$, $Bernouli(\theta_i)$ for short. 
A sufficient test statistic for testing the hypothesis in \eqref{testproblem} is $T(\pmb{X})=\sum_{r=1}^n X_r$ which is distributed as a Binomial random variable with parameters $n$ and $\theta_i$, $i=1,2$ and we shall denote this by $Bin(n,\theta_i).$ The respective $p$-values with their CDFs are calculated using Equations 
\eqref{eq:ump_p_value},\eqref{eq:ump_p_value_cdf},\eqref{eq:rand2_p_value}, and \eqref{eq:rand2_p_value_cdf}.
The critical constants $C_n$ and $D_n$ are given by
$C_n=F^{-1}_{Bin(n, \theta_1)}(1-t)$ and $D_n=F^{-1}_{Bin(n, \theta_2)}(t)$ for $t\in (0,1)$ where $F^{-1}(\bullet)$ denotes the quantile of a binomial random variable with parameters $n$ and $\theta.$ The randomization constants $\gamma_n$ and $\delta_n$ for large sample sizes and for arbitrary $t\in (0,1)$ are given by $\gamma_n=\{F_{Bin(n,\theta_1)}(C_n)-(1-t)\}\{f_{Bin(n,\theta_1)}(C_n)\}^{-1}$ and $\delta_n=\{t-F_{Bin(n,\theta_2)}(D_n-1)\}\{f_{Bin(n,\theta_2)}(D_n)\}^{-1},$ where $F_{Bin(n,\theta)}$ denotes the CDF and $f_{Bin(n,\theta)}$ the probability mass function of binomial variable with parameters $n$ and $\theta$. 
\end{example}

In this section,  as mentioned before, we consider the individual test problem where $k=1$. We are interested in finding if randomization is beneficial when the equivalence limit $\Delta$ increases or decreases and if the power functions for the $p$-values are monotonic in sample size. Furthermore, we seek to find if the level of conservativity of the $p$-values depends on the sample sizes.

\subsection{Sample size versus power}

\begin{figure}[H]
\begin{center}
\includegraphics[width=5in]{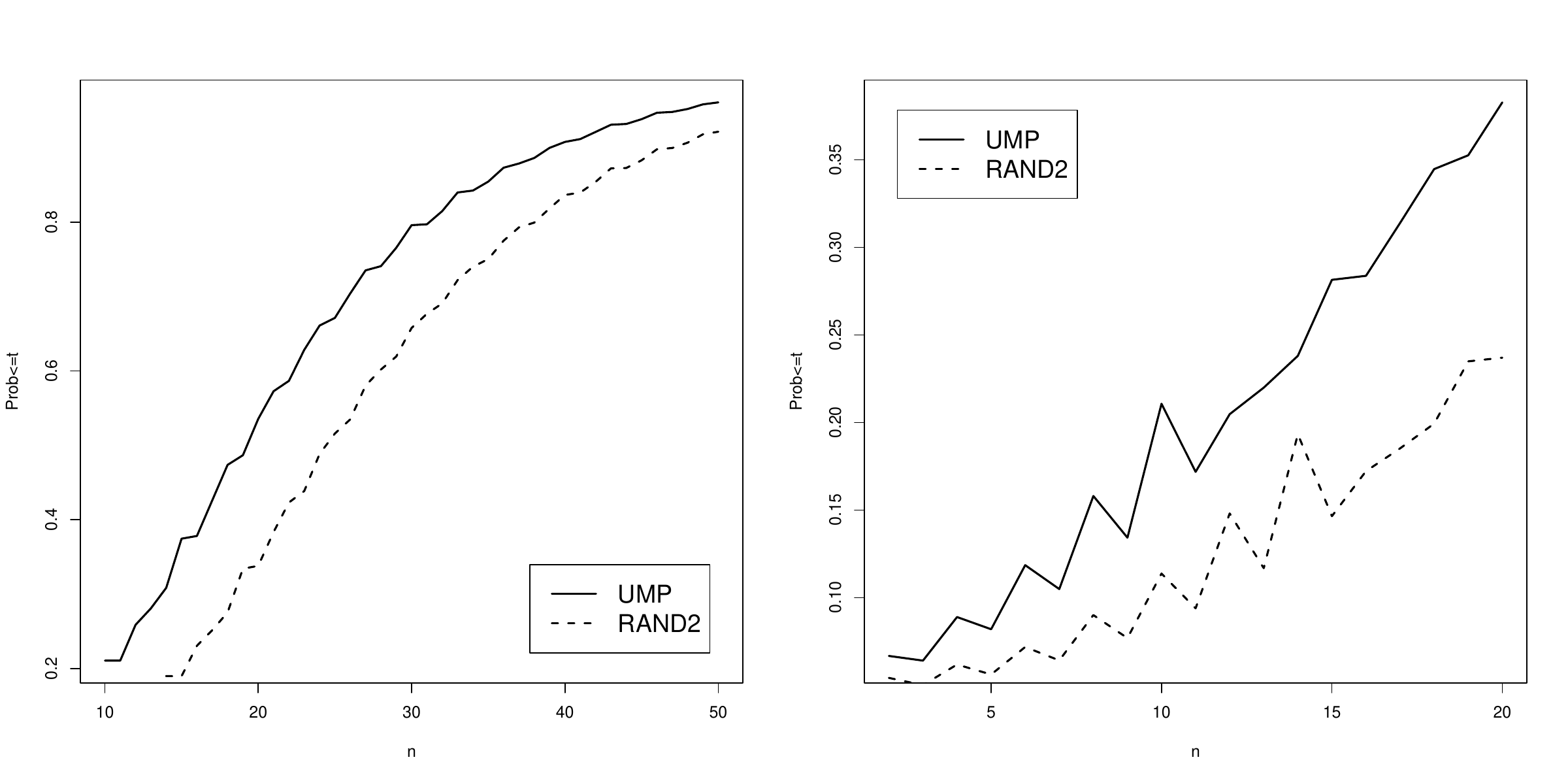}
\caption{The power function for the UMP and RAND2 $p$-values against different sample sizes for $c=0.5$, $\theta_1=0.25$, and $\theta_2=0.75$. Furthermore, we set $\theta=0.5$ in the left panel and $\theta=0.4$ in the right.}
\label{sample_power}
\end{center}
\end{figure}

We expect that the power function for a test would be strictly increasing with an increase in sample size. A power function that is strictly increasing with an increase in the sample size is ideal for sample size planning since an additional observation cannot lower the power. In the case of discrete models, \cite{finner2001increasing} showed that it is possible for the power of the (least favorable configuration) LFC-based $p$-value at a sample of size $n$ to coincide over the entire parameter space with that of size $n+i$, for small $i\in \mathbb{N}$. We illustrate in the second panel of Figure \ref{sample_power} and for the model in Example \eqref{binomial_model} that this paradoxical behavior can also occur for the UMP $p$-value and cannot be corrected even by use of randomization. The problem occurs for small samples with the chosen true parameter $\theta$ too close to the boundary of the alternative hypothesis. To generate Figure \ref{sample_power}, we set the tuning parameter $c=0.5$, $\theta_1=0.25$, and $\theta_2=0.75$ in both panels. Furthermore, we choose $\theta=0.5$ as the true parameter under the alternative hypothesis in the left panel and $\theta=0.4$ in the right.

\begin{figure}[H]
\begin{center}
\includegraphics[width=3 in]{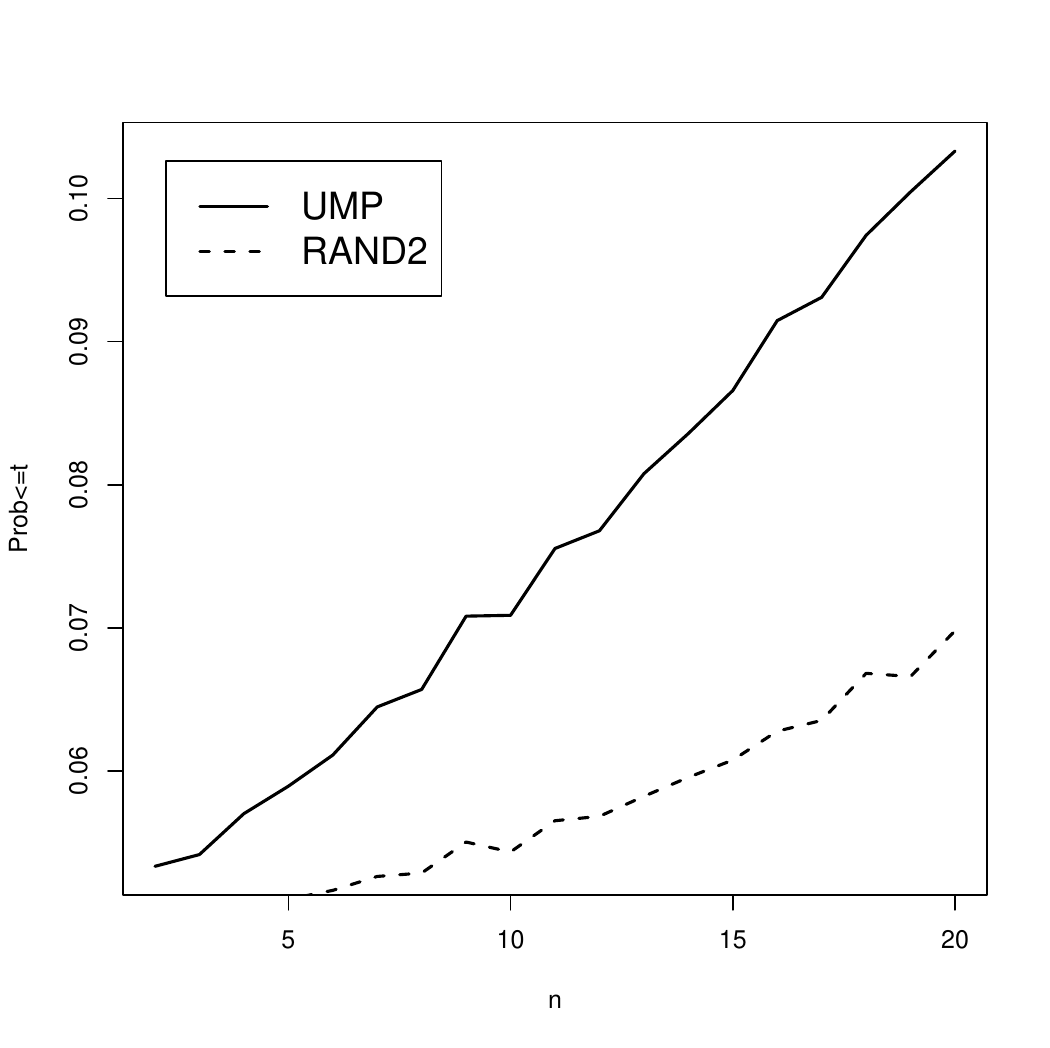}
\caption{The power function for the UMP and RAND2 $p$-values against different sample sizes for $c=0.5$, $\theta_1=0.35$, and $\theta_2=0.75$. We maintain the true parameter under the alternative hypothesis at $\theta=0.4$.}
\label{sample_power2}
\end{center}
\end{figure}

On the left panel in Figure \ref{sample_power}, both power functions are strictly increasing with an increase in the sample size. On the right panel, both power functions are not monotonically increasing with larger sample sizes. We further illustrate in Figure \ref{sample_power2} that this paradoxical behavior of the power function of the UMP $p$-value in the right panel of Figure \ref{sample_power} does not occur for small equivalence limit  $\Delta$. To generate Figure \ref{sample_power2}, we maintain the parameter settings as in the right panel of Figure \ref{sample_power} but only change $\theta_1$ to $0.35$ so that the resulting $\Delta$ is decreased compared to the initial one. 

From Figure \ref{sample_power2}, the power functions for the UMP and RAND2 $p$-values are now strictly increasing with an increase in the sample size for most $n$. The problem of the power function failing to be strictly increasing with an increase in the sample size is partially dealt with, though not completely removed. Shrinking $\Delta$ from both sides, however, worsens the problem in the right panel of Figure \ref{sample_power}. Finally, we provide Theorem \eqref{theorem:4.1} with a proof in the appendix to further justify the claims in the right panel of Figure \eqref{sample_power}.

\begin{theorem}[ Monotonicity of the power functions]
The CDFs of the UMP and RAND2 $p$-values are strictly increasing with an increase in the sample size $n$ for any fixed parameter value $\theta$ under the alternative hypothesis. Consequently, for any significance level and a fixed parameter value $\theta$ under the alternative hypothesis, the power of the corresponding test is monotonically increasing with an increase in the sample size $n$.
\label{theorem:4.1}
\end{theorem}

\subsection{Conservativity of the $p$-values}

\begin{figure}[H]
\begin{center}
\includegraphics[width=5.2 in]{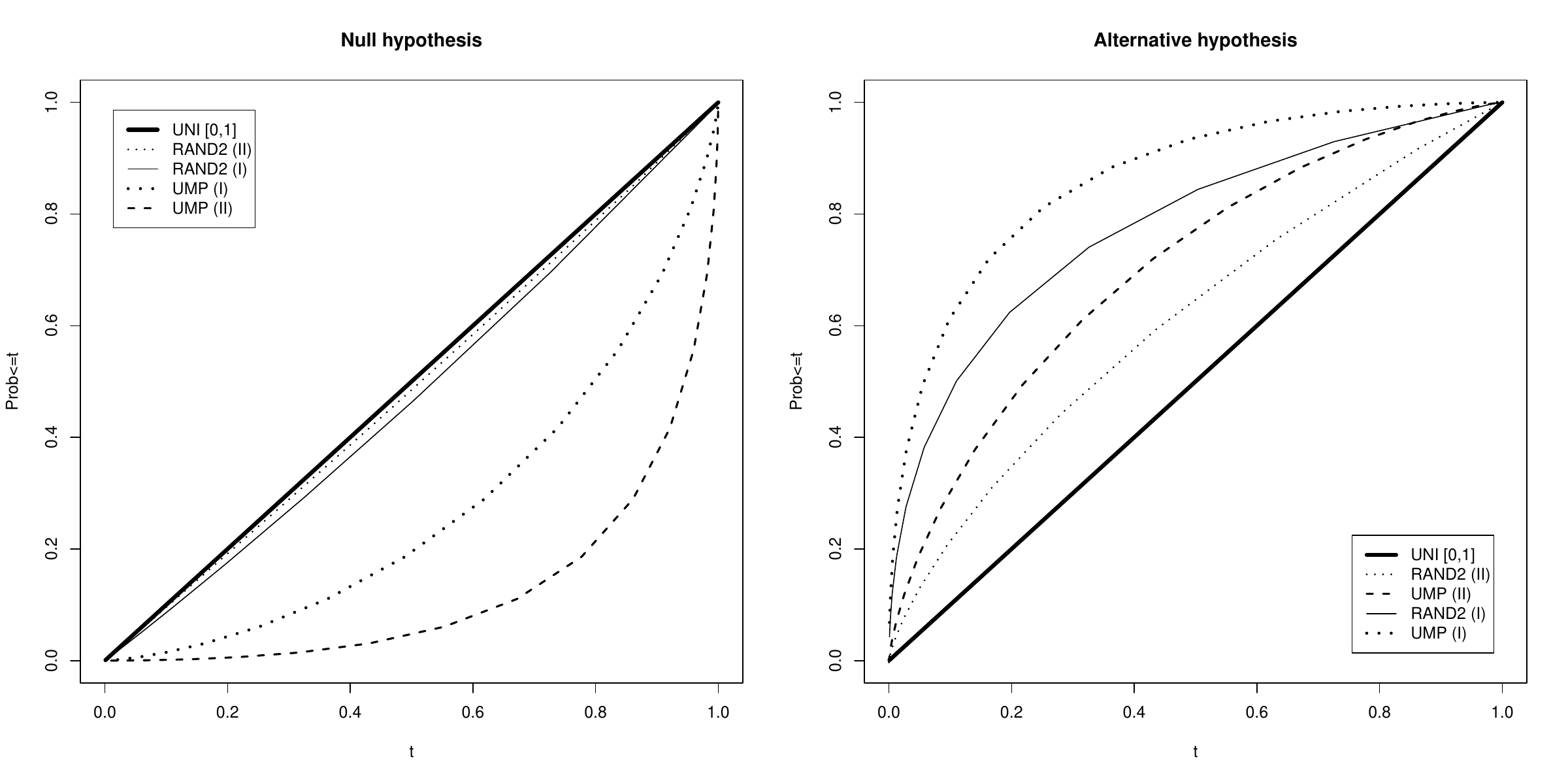}
\caption{The CDFs of the UMP and RAND2 $p$-values against $t$ for $n=50$ and $c=0.5$. We choose the true parameter  $\theta=0.2$ under the null hypothesis and $\theta=0.35$ under the alternative hypothesis. Furthermore, we set $\theta_1=0.25$, and $\theta_2=0.75$ in the first case (I) and $\theta_1=0.3$, and $\theta_2=0.75$ in the second case (II).}
\label{conservativeness1}
\end{center}
\end{figure}

As mentioned in the introduction, we expect that the distribution of a $p$-value under the null hypothesis is close to that of a $UNI(0,1)$ distribution. A $p$-value can fail to meet this requirement and hence be conservative, meaning it is stochastically greater than the uniform distribution. We illustrate for the model in Example \ref{binomial_model} that among the two $p$-values, only RAND2 $p$-value comes close to meeting this requirement and is therefore less conservative than the UMP $p$-value. We illustrate in Figure \ref{conservativeness1} that utilizing the two-stage randomized $p$-value reduces the conservativeness of the UMP $p$-value. In this figure, we consider two cases where we have set $\theta_1=0.25$ and $\theta_2=0.75$ in the first case and  $\theta_1=0.3$ and $\theta_2=0.75$ in the second case. For both cases, we use a sample of size $n=50$ and set the tuning parameter to $c=0.5$. The chosen parameter $\theta$ is $0.2$ under the null and $0.35$ under the alternative hypothesis. Notice that the equivalence limit $\Delta$ in the first case is larger than the second case. The reason for using these two equivalence limits is to find if the $p$-values will become more or less powerful (or conservative) depending on the size of the equivalence limit $\Delta$.

From Figure \ref{conservativeness1}, the CDF of the UMP $p$-value under the null hypothesis is far from the $UNI(0,1)$ line compared to the one for RAND2 $p$-value in both cases. Therefore, the UMP $p$-value is more conservative compared to RAND2 $p$-value. Under the alternative hypothesis, the CDF of the UMP $p$-value is also far from the $UNI(0,1)$ line compared to the one for RAND2 $p$-value in both cases. Therefore, as expected, the power of the UMP $p$-value exceeds that of RAND2 $p$-value. This power loss is the price for using our randomized $p$-value. We can use conditioning (cf. \cite{zhao2019multiple})  to improve the power of tests based on these $p$-values.

Under the same parameter configurations and only shrinking the equivalence limit $\Delta$, the UMP $p$-value becomes less powerful and more conservative. The two-stage randomized $p$-value also becomes less powerful, but the conservativeness of the $p$-value reduces even further. Notice that we shrink the equivalence limit by increasing $\theta_1$ while holding $\theta_2$ constant. Since the chosen parameter $\theta$ under the null hypothesis is also constant, this parameter will now be too far from the boundary of the resulting equivalence limit. Shrinking the equivalence limit by increasing $\theta_1$ and reducing $\theta_2$ lowers the power but does not affect the level of conservativeness for both $p$-values. Furthermore, holding $\theta_1$ constant and reducing the equivalence limit by decreasing $\theta_2$ does not affect both the power and the level of conservativeness for both the $p$-values. These observations on the CDf for the two $p$-values under the null (alternative) for the model in Example \ref{binomial_model} depends on whether the chosen parameter under the null (alternative) is such that $\theta\leq \theta_1$ or  $\theta\geq \theta_2$ ($\theta<0.5$ or $\theta>0.5$) and we cannot provide a general statement.

A similar trend in Figure \ref{conservativeness1} occurs when the equivalence limit is kept constant with the chosen true parameter under the null too far from the null boundary or the one under the alternative too close to the boundary. Furthermore, the same behavior in Figure \ref{conservativeness1} occurs when the chosen true parameter under the null or alternative hypothesis is held constant and $\Delta$ is shifted by an $\epsilon\in \mathbb{R}$ so that the new interval is of the form $[\theta_1+\epsilon,\theta_2+\epsilon].$ 

\begin{figure}[H]
\begin{center}
\includegraphics[width=5.2 in]{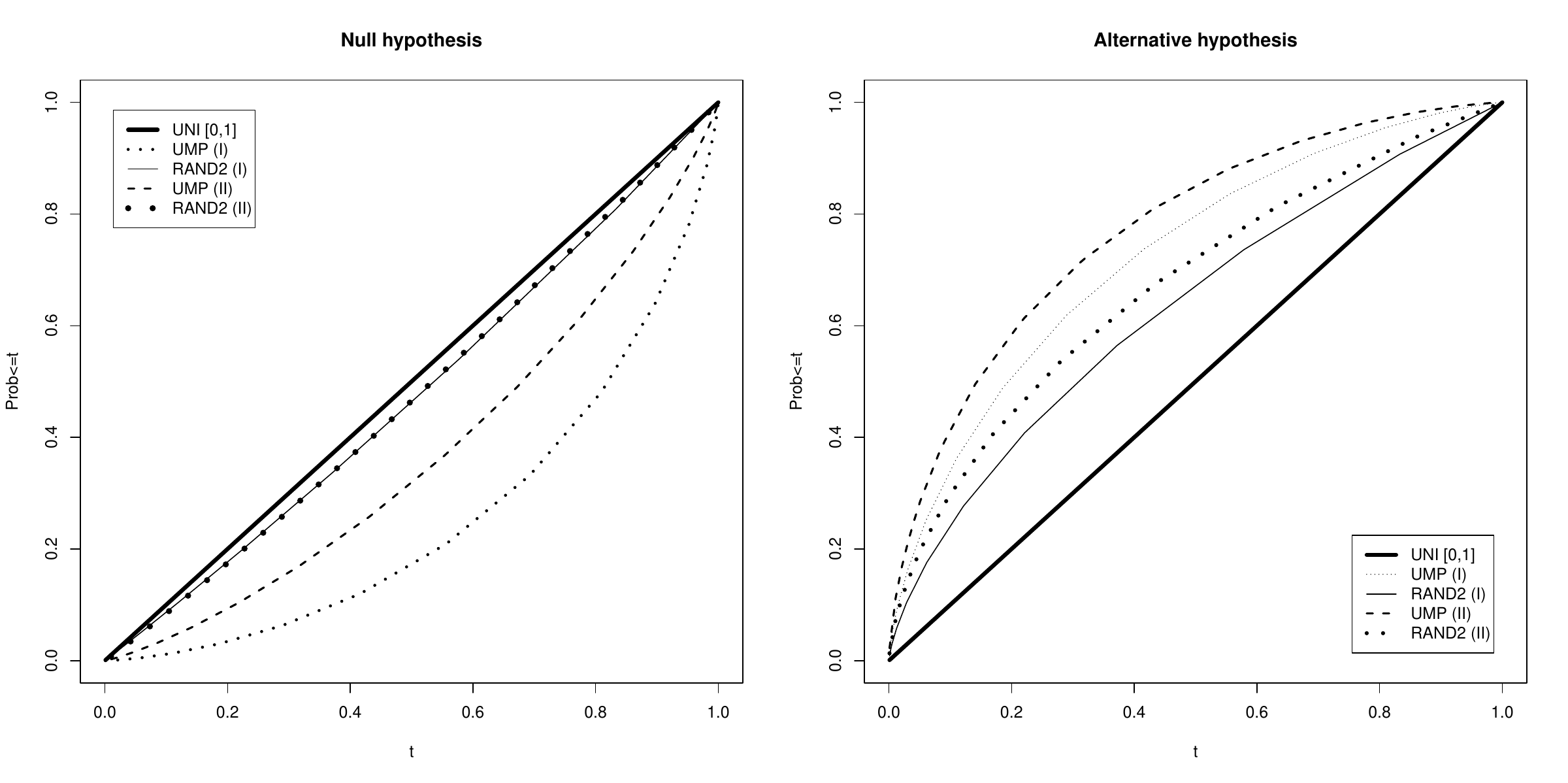}
\caption{The CDFs of the UMP and RAND2 $p$-values against $t$ for $n=50$ and $c=0.5$. Furthermore, we set $\theta_1=0.2$, $\theta_2=0.7$, and the chosen true parameters are $\theta=0.15$ under the null and $\theta=0.25$ under the alternative hypothesis in the first case (I). In the second case (II), we set $\theta_1=0.3$, $\theta_2=0.8$, and the chosen true parameters are $\theta=0.27$ under the null and $\theta=0.37$ under the alternative hypothesis.}
\label{conservativeness3}
\end{center}
\end{figure}

Shifting the equivalence limit and the chosen parameter under the null or alternative hypothesis with an $\epsilon$ leads to different behaviors for the CDFs. We illustrate this in Figure \ref{conservativeness3} using $n=50$ and the tuning parameter set at $c=0.5$. 
Furthermore, we consider two cases where in the first one, we set $\theta_1=0.2$, $\theta_2=0.7$, and the chosen true parameter $\theta=0.15$ under the null and $\theta=0.25$ under the alternative hypothesis. In the second case, we shift the parameters by an $\epsilon_1=0.1$ so that $\theta_1=0.3$ and $\theta_2=0.8$. The true parameters are shifted by $\epsilon_2=0.12$ so that $\theta=0.27$ under the null and $\theta=0.37$ under the alternative hypothesis. Notice that $\epsilon_2>\epsilon_1.$

\begin{figure}[H]
\begin{center}
\includegraphics[width=3in]{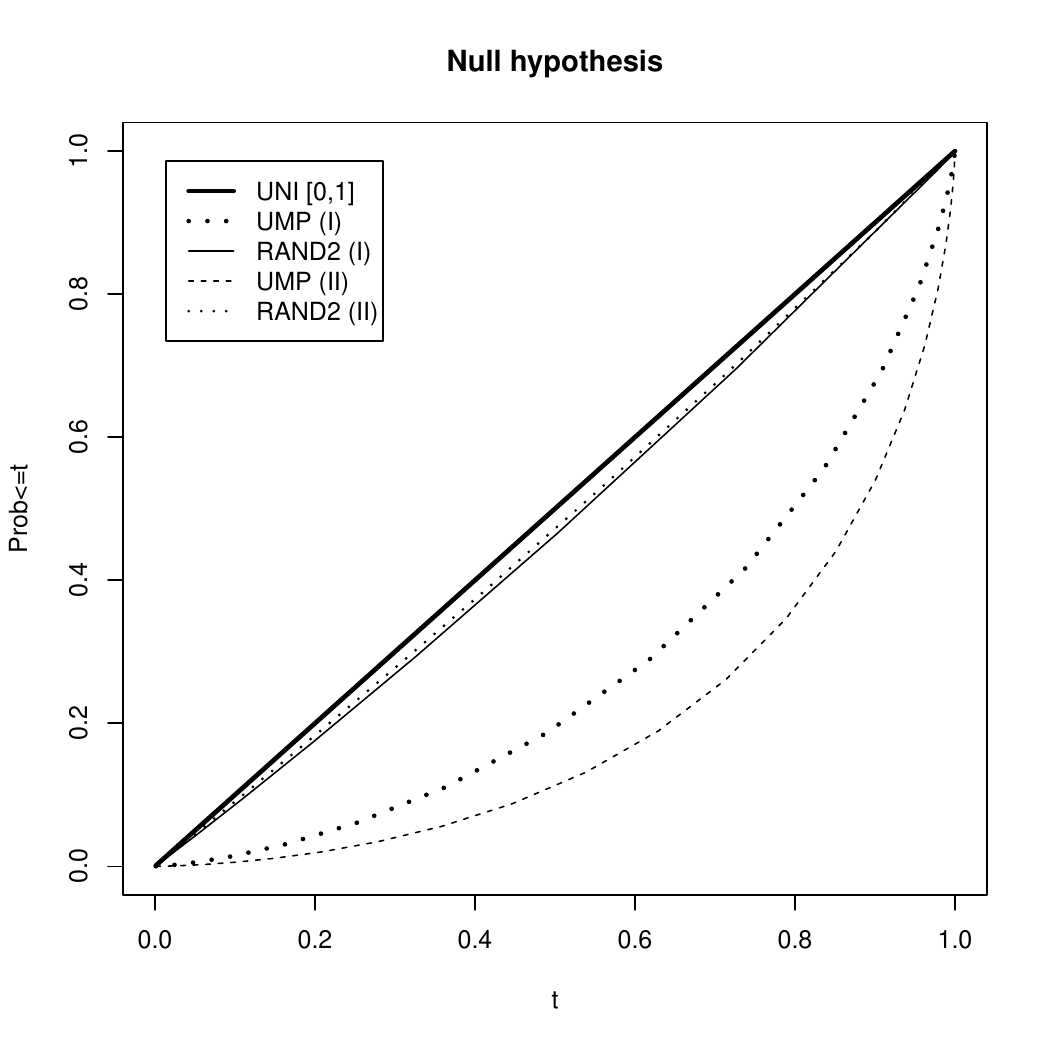}
\caption{The CDFs of the UMP and RAND2 $p$-values against different values of $t$ with $c=0.5$, $\theta_1=0.25$, $\theta_2=0.75$, and with $\theta=0.2$ as the chosen parameter under the null hypothesis. Furthermore, we use $n=50$ in the first case (I) and $n=100$ in the second case (II).}
\label{conservativeness2}
\end{center}
\end{figure}

From Figure \ref{conservativeness3} and with the parameters shifted as described, the CDF for the UMP $p$-value under the null hypothesis moves closer to the $UNI(0,1)$ line while there is no change in the one for RAND2 $p$-value. The CDFs for both $p$-values under the alternative hypothesis move away from the $UNI(0,1)$ line. These results hold true for any $\epsilon_1$ and $\epsilon_2$ as long as $\epsilon_2>\epsilon_1$. For $\epsilon_2\leq \epsilon_1$, the CDFs for both the $p$-values under the null and alternative hypothesis behave exactly as in Figure \ref{conservativeness1}. Next, we give Figure \ref{conservativeness2} to illustrate the behavior of the CDFs for the two $p$-values under the null hypothesis using the same parameter configurations as in Figure \ref{conservativeness1} except that the sample size $n$ is not constant. Again, we consider two cases but with $n=50$ in the first case (I) and $n=100$ in the second case (II).

From Figure $\ref{conservativeness2}$, the CDF of the UMP $p$-value moves away while the one for RAND2 $p$-value moves closer to the $UNI(0,1)$ line as sample size increases. Therefore, the UMP $p$ value becomes more conservative while the RAND2 $p$ value is less conservative as the sample size increases.

\subsection{Maximum power}

\begin{figure}[H]
\begin{center}
\includegraphics[width=5.3 in]{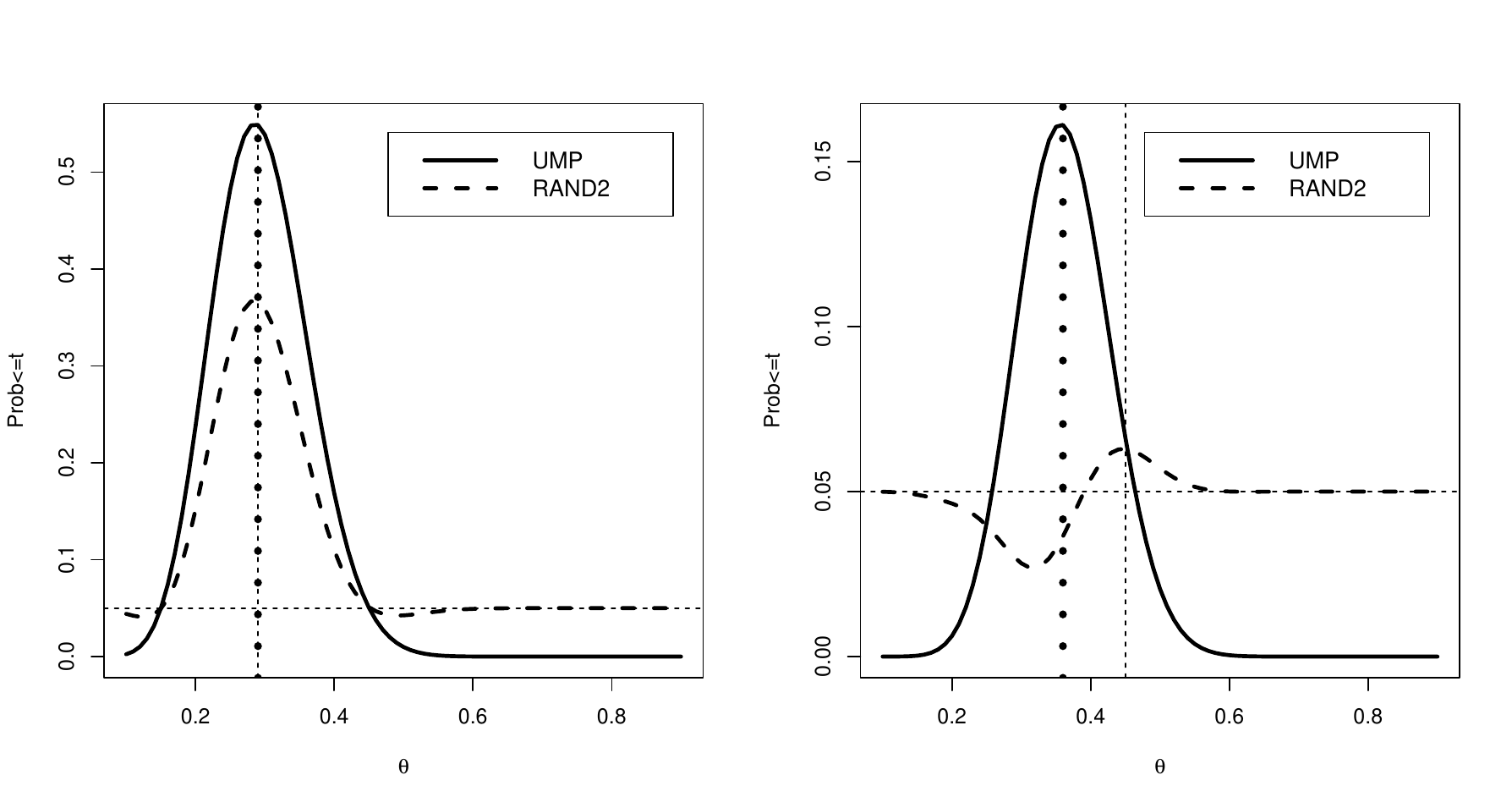}
\caption{The CDFs for the UMP and RAND2 $p$-values against the chosen parameter $\theta$ for $c=0.5$ and $n=50$. We set $\theta_1=0.15$ and $\theta_2=0.45$ in the left panel and  $\theta_1=0.25$ and $\theta_2=0.45$ in the right one. The vertical lines intersect the respective CDF curves at their maximum and the $x$ axis at the parameter value that maximizes those CDFs. The bold vertical line is for the UMP $p$-value while the thin dotted line is for RAND2 $p$-value. Furthermore, the thin dotted horizontal lines intersect the $y$ axis at the value of $\alpha$.}
\label{max_power}
\end{center}
\end{figure}

We find the parameter value that maximizes the CDF of the two $p$-values under the alternative hypothesis for a given equivalence limit $\Delta$. Once we get this parameter, we can choose it as our parameter under the alternative hypothesis, so we always get the maximum power. Furthermore, one may wonder if the value of this parameter depends on $\Delta$ or if two or more such parameters exist within the alternative parameter space. We generate Figure \ref{max_power} to address these questions for Example \ref{binomial_model}, where we have set $c=0.5$ and used $n=50$ as our sample size. Furthermore, we use $\theta_1=0.15$ and $\theta_2=0.45$ in the left panel of Figure \ref{max_power} and $\theta_1=0.25$ and $\theta_2=0.45$ in the right one.

\begin{figure}[H]
\begin{center}
\includegraphics[width=5.2 in]{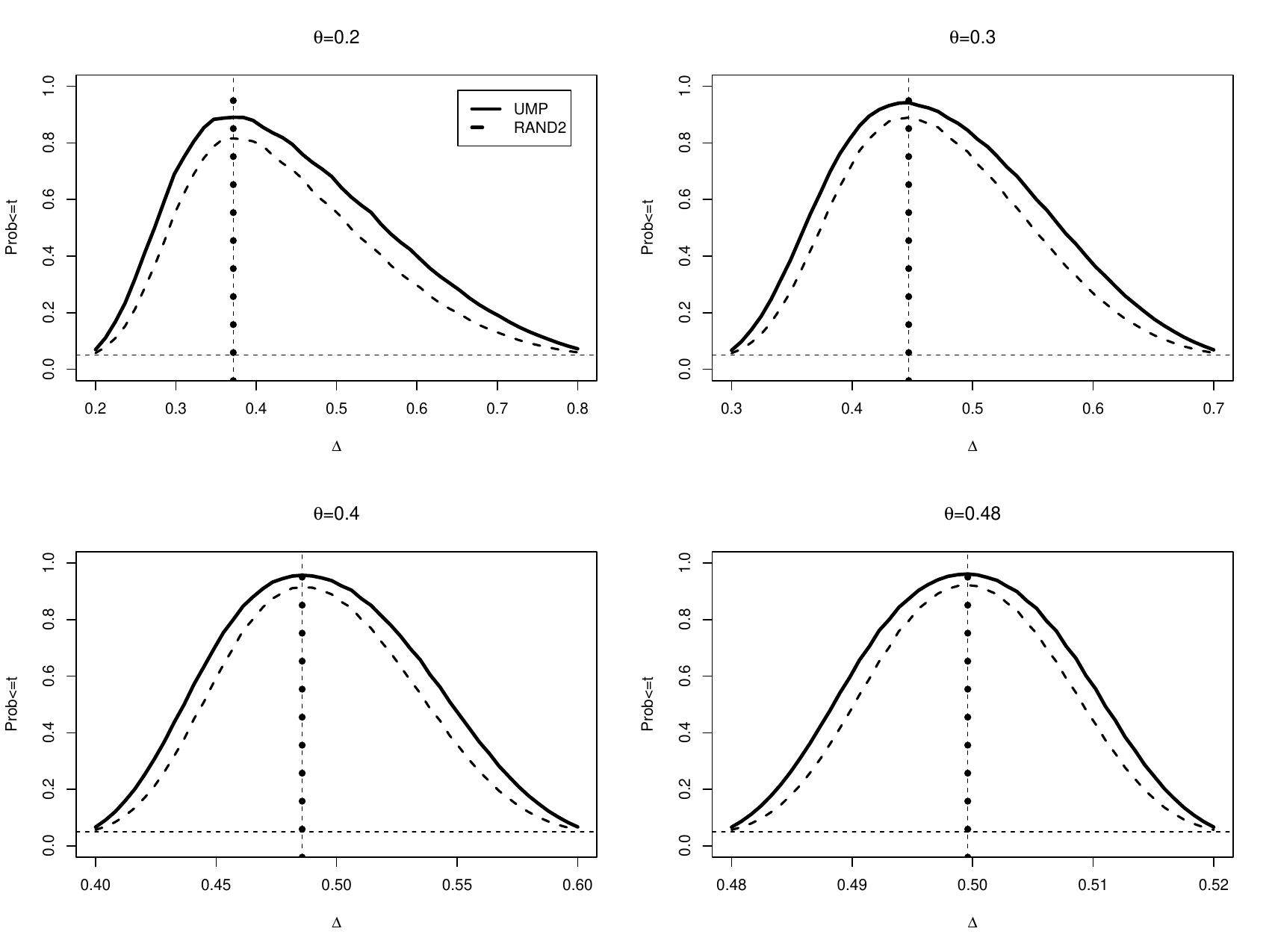}
\caption{The CDF under the alternative hypothesis for the UMP and RAND2 $p$-values against the equivalence limit $\Delta$ for $c=0.5$ and $n=50$. The chosen parameters under the alternative hypothesis are $\theta=0.2, 0.3,0.4,$ and $0.48$, respectively, from left to right. The vertical lines intersect the respective CDF curves at their maximum and the $x$ axis at the $\Delta$ value, which maximizes those CDFs. The bold vertical line is for the UMP $p$-value while the thin dotted line is for RAND2 $p$-value. Furthermore, the thin dotted horizontal lines intersect the $y$ axis at the value of $\alpha$.}
\label{delta_max_power}
\end{center}
\end{figure}

From Figure \ref{max_power} and for a large equivalence limit $\Delta$ like the one in the left panel, the maximum of the CDF under the alternative hypothesis for the two $p$-values always occurs at the midpoint of the interval $\Delta$. For a small $\Delta$ like the one in the right panel, the maximum of the CDF under the alternative hypothesis for RAND2 $p$-value can occur at a point too close to $\theta_1$ or $\theta_2$. The one for the UMP $p$-value occurs at the midpoint throughout, and it does not matter how small $\Delta$ becomes. Also, for both $p$-values, only a single parameter value maximizes the CDF under the alternative hypothesis. Moreover, the behavior of the CDFs in the right panel further confirms that RAND2 $p$-value, unlike the UMP $p$-value, is not unbiased. To conclude this section, we give a figure illustrating the power for the two $p$-values against the equivalence limit $\Delta$. To generate Figure \eqref{delta_max_power}, we set $c=0.5$, $n=50$, and choose $\theta=0.2,0.3,0.4,$ and $0.48$ as the true parameters under the alternative hypothesis. Moreover, we use different values of $\theta_1$ and $\theta_2$  to get different equivalence limits since $\Delta=\theta_2-\theta_1$.

From Figure \ref{delta_max_power}, as is expected, the range of $\Delta$ in each panel is from the chosen parameter value $\theta$ under the alternative hypothesis to $1-\theta$. For example, in the first panel, the parameter is $\theta=0.2$ and $\Delta$ ranges from $\theta=0.2$ to $1-\theta=0.8$. The value of $\Delta$ that gives the maximum power $\Delta_{max}$ for both the $p$-values corresponds to the value of the chosen parameter under the alternative hypothesis $\theta_{max}$, which also gives the maximum power. The maximum CDFs for the two $p$-values move closer to $\Delta_{max}$ as the chosen parameter under the alternative hypothesis moves closer to $\theta_{max}$.

\section{Estimation of the proportion of true null hypotheses}\label{multiple_interval}

\subsection{Introduction}

In this section, we extend our discussions from Section \ref{interval_composite} to the case when $k>1$ hypotheses are of interest. Conducting these hypotheses at level $\alpha$ increases the probability of type I errors since we do not account for the multiplicity of the problem. It is therefore important to account for this multiplicity by doing, for example, a familywise error rate (FWER) control. One commonly used method for familywise error control at level $\alpha$ is the Bonferroni adjustment (cf. \cite{bonferroni1936teoria}). The Bonferroni procedure adjusts the raw $p$-values $p_1,\ldots,p_k$ by multiplying them by the number of hypotheses $k$. We reject the null hypothesis if an adjusted $p$-value is less than or equal to $\alpha$. The Bonferroni procedure guarantees that the FWER is at most $\alpha$ regardless of the ordering or the dependence structure of the $p$-values. The Bonferroni procedure can be conservative when large proportions of null hypotheses are false. The adjustment also maintains FWER at levels below $\pi_0 \alpha$ instead of $\alpha$ where $\pi_0=k_0/k$ is the proportion of true null hypotheses. When the true number of null hypotheses $k_0<k$, the individual tests are conducted at a higher level $\alpha/k_0$ instead of $\alpha/k$, leading to a higher power for the testing procedure. We refer to this as the adaptive Bonferroni procedure, ABON for short. Since in practice we never really know the number (proportion) $k_0$ ($\pi_0$),  we make use of ABON combined with the plug-in (ABON+plug-in) procedure of \cite{finner2009controlling} to estimate $\pi_0$. The ABON+plug-in procedure, unlike closed testing procedures (like \cite{hommel1988stagewise} and \cite{hochberg1988sharper}), provides a theoretical guarantee to control the type I error rate at the desired level. One classical but still commonly used estimator for $k_0$ is the \cite{schweder1982plots} estimator. It is given by 
\begin{equation}\label{schweder-def}
\hat{k}_0\equiv \hat{k}_0(\lambda)= k \cdot  \frac{1-\hat{F}_k(\lambda)}{1-\lambda},
\end{equation}
where $\lambda\in [0,1)$ is a tuning parameter and $\hat{F}_k$ is the empirical CDF (ecdf) of the $k$ marginal $p$-values. It is often suggested in practice to choose $\lambda=0.5$. 
One crucial prerequisite for the applicability of this estimator is that the marginal $p$-values $p_1, \ldots, p_k$ are (approximately) uniformly distributed on $(0,1)$ under the null hypothesis; see, e.\ g., \cite{dickhaus2013randomized}, \cite{hoang2021usage} and the references therein for details.
The randomized $p$-values considered in this work are close to meeting the uniformity assumption, whereas the non-randomized $p$-values are over-conservative when testing two one-sided composite null hypotheses, especially in discrete models. Typically, the estimated value of $k_0$ becomes too large if many null $p$-values are conservative and the estimator from \eqref{schweder-def} is employed.  

\subsection{Empirical distributions}\label{empirical}

\begin{figure}[H]
\begin{center}
\includegraphics[width=4in]{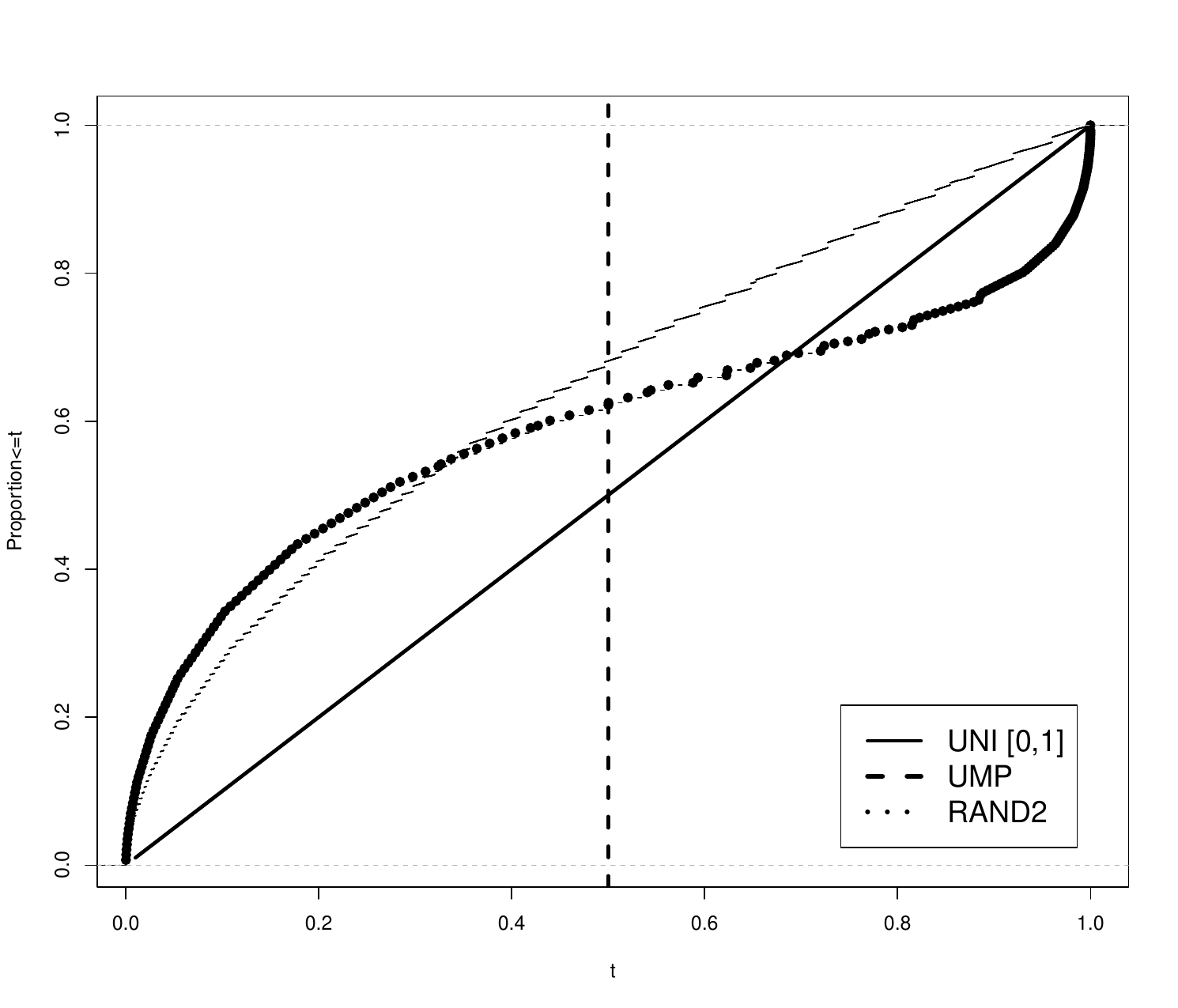}
\caption{Empirical CDF of the UMP $p$-value (black curve) and the two-stage randomized (RAND2) $p$-value (grey curve) for $k=1000$, $\lambda=0.5$, $c=0.5$, and $\pi_0=0.7$. We set $\theta_1=0.25$ and $\theta_2=0.75$. Furthermore, we choose the true parameter under the null as $\theta=0.18$ 
and otherwise $\theta=0.37$. The dashed vertical line intersects the $x$ axis at the value of $\lambda$.}
\label{figure_3_ecdf}
\end{center}
\end{figure}

To illustrate the implication of using our proposed two-stage randomized $p$-value in multiple testing, we employ a graphical algorithm in computing $\pi_0$. This algorithm connects the points $(\lambda,\hat{F}_k)$ with the point $(1,1)$.  We draw a straight line to connect the two points and extend this line to intersect the $y$ axis at the point $1-\hat{\pi}_0$. The best $p$-value for use in the estimation of $\pi_0$ is that for which the resulting straight line meets the $y$ axis at a point that is very close to the actual $1-\pi_0$. We require the empirical CDF of the $p$-value not to lie below the $UNI(0,1)$ line for this to be actualized. 
Another way to look at this is to find the gradient of the resulting straight line between the points $(\lambda, \hat{F}_k)$ and $(1,1)$, and this should give you an estimate of $\pi_0$. If the resulting ECDF line lies below the $UNI(0,1)$ line between those two points, then it has to be a curve whose gradient can only be at a tangent and hence will give a poor estimate of $\pi_0.$ 
To generate Figure \ref{figure_3_ecdf}, we let the number of hypotheses to be $k=1000$, the tuning parameters $c$ and $\lambda$ are both set at $0.5$, and use a sample of size $n=50$. We take the proportion of true null hypotheses to be $\pi_0=0.7$ and set $\theta_1=0.25$ and $\theta_2=0.75$. Furthermore, to calculate the UMP-based $p$-value, the parameter $\theta$ under the null and alternative hypothesis are chosen as $0.18$ and $0.37$, respectively.

From Figure \ref{figure_3_ecdf}, RAND2 $p$-value outperforms the UMP $p$-value since its ECDF lies above the $UNI(0,1)$ line for all values of $t\in(0,1)$. Furthermore, an extension of a straight line from the points $(1,1)$ to $(\lambda, F_k^{RAND2})$ as earlier mentioned, meets the $y$ axis at a point which is close to $1-\pi_0$. 

\subsection{Simulation study}

We now conduct a simulation study based on real-world data to support the claim in Section \ref{empirical} that RAND2 $p$-value outperforms the UMP $p$-value in estimating the proportion of true null hypotheses in multiple testing. We use the publicly available Coronavirus Disease 2019 (COVID-19) data taken from  \url{https://github.com/CSSEGISandData/COVID-19} (cf. \cite{dong2020interactive}). It consists of confirmed COVID-19 cases and recoveries for the United States of America as of $12^{th}$ May 2020. The data set has $k=58$ regions. After cleaning the data by removing all the missing values, we have $k=47$ regions for our analysis. We select an interval of recovery rates $\theta_1$ and $\theta_2$ and conduct a TOST to find if the true rates from the data set belong to these intervals. We use a Monte Carlo simulation to assess the (average) performance of the UMP and RAND2 $p$-values in estimating $k_0$. We set the constant $c$ and the tuning parameter $\lambda$ in \eqref{schweder-def} to  $0.5$ for all the simulations. The recovery rates from the data set are assumed to be the true proportions. Using these rates and the number of confirmed cases, we generate a new data set on the computer for calculating the $p$-values. Using different values of $\theta_1$ and $\theta_2$, we obtain various values of $k_0 \in \{0, \ldots, 47\}$ depending on $\Delta$. Since we are utilizing randomized $p$-values in \eqref{schweder-def}, we average the estimated value of $k_0$ over the $10{,}000$ Monte Carlo repetitions. For exemplary purposes, we present ten choices of $\theta_1$ and $\theta_2$ in Table $\ref{tabletwo}$. A detailed description of the simulation is provided in Algorithm \ref{algo-sec5}.

\begin{algorithm}
\caption{Computation of the proportion of true null hypotheses}\label{algo-sec5} 
\begin{enumerate}
\item[1)] For each of the $k=47$ regions in the COVID-19 data set, find the proportions of recoveries $\theta_i,\ i=\{1,\ldots,47\}$ and use these as the assumed true proportions (i.\ e., as the assumed ground truth) in the simulations.

\item[2)]For each $\theta_i$  from step 1.) and for each of the sample sizes $n_i$ for $ \ i=\{1,\ldots,47\}$, simulate a single data point $x_i$ from $Bin(n_i,\theta_i)$.  

\item[3)]Select two proportions  $\theta_1, \theta_2\in[0,1]$ such that $\theta_1<\theta_2$ as the null values to be tested against. Take $k_0$ as the number of values $i \in \{1, \ldots, k\}$ fulfilling that $\theta_i\leq \theta_1$ or $\theta_i\geq \theta_2$. We use the selected $\theta_1, \theta_2$ as well as the numbers $x_i$, and $n_i$ from step 2.) in the computation of the $p$-values, where $i\in \{1, \ldots, k\}$. This step generates a pair of $p$-values for the UMP $p$-value since we decompose the null hypothesis in \eqref{testproblem} into a lower- and an upper-sided test. Denote these $p$-values by $p_l$ and $p_u$, respectively. In each case, pick $\text {max}\ (p_l,p_u)$ which is the maximum of the two $p$-values.

\item[4)]Compute the statistic in Equation \eqref{schweder-def} $r={10,000}$ times for the UMP and RAND2 $p$-values and take the mean.  
\end{enumerate}
\end{algorithm}

The results from our simulation study based on the Algorithm \ref{algo-sec5} are presented in Table \ref{tabletwo}.

\begin{table}[h]
\caption{Estimates of the number of true null hypotheses.}\label{tabletwo}%
\begin{tabular}{@{}llllll@{}}
\toprule
$\theta_1$ &$\theta_2$& $\Delta$&$k_0$ & $\hat{k}^{UMP}_0$ & $\hat{k}^{RAND2}_0$\\ 
\midrule
0.4791& 0.5413& 0.0622& 45& 90.0050& 44.3586\\
0.4509& 0.5681& 0.1173& 43& 86.0006& 43.1554\\
0.4444& 0.5946& 0.1502& 40& 80.0034& 39.4800\\
0.4066& 0.6800& 0.2734& 34& 67.9996& 33.5392\\
0.3389& 0.7219& 0.3830& 31& 60.0002& 33.7460\\
0.3188& 0.7478& 0.4290& 29& 55.9958& 28.2846\\
0.3076& 0.7566& 0.4491& 28& 55.9958& 28.6418\\
0.2963& 0.9029& 0.6065& 16& 32.0070& 15.2562\\
0.2725& 0.9319& 0.6594& 12& 26.0192& 12.9908\\
0.2456& 0.9399& 0.6942& 12& 24.6468& 13.9496\\
\botrule
\end{tabular}
\end{table}

From Table \ref{tabletwo}, for whatever value of $\Delta$, RAND2 $p$-value outperforms the UMP $p$-value by giving estimates which are on average close to the actual value of $k_0$. Also, as is expected, the number of true null hypotheses $k_0$ decreases with an increase in the interval $\Delta$.

\subsection{Role of the tuning parameter $\lambda$}

\begin{figure}[H]
\begin{center}
\includegraphics[width=5in]{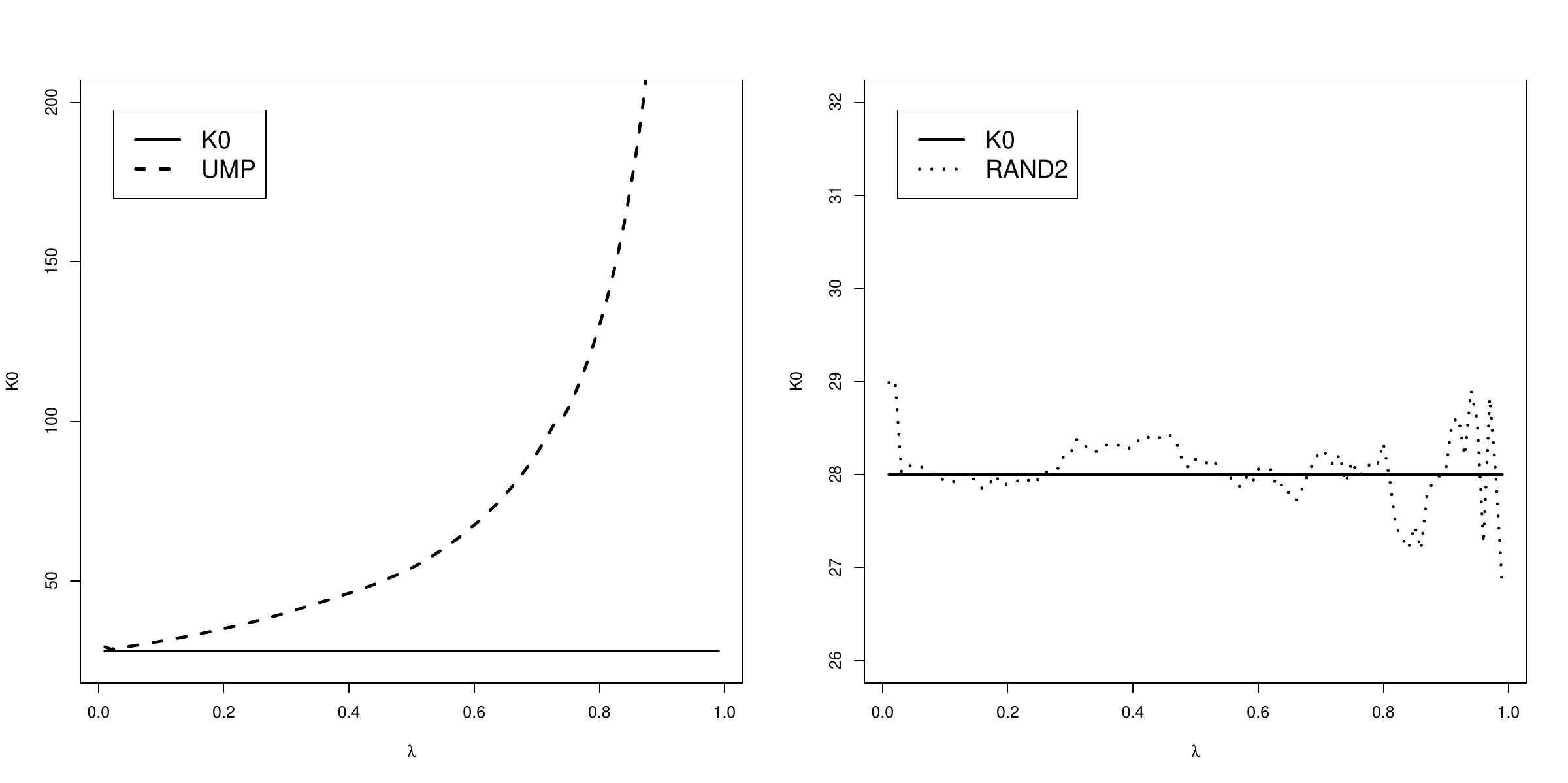}
\caption{An illustration of the number of true null hypotheses versus $\lambda$ for the UMP and the two-stage randomized (RAND2) $p$-values for $\theta_1=0.2963$, $\theta_2=0.7566$, and $c=0.5$.}
\label{f:role_of_lambda}
\end{center}
\end{figure}

In this section, we investigate the role of the tuning parameter $\lambda$ in the estimator given in \eqref{schweder-def}  when using the two $p$-values. Proper choice of this parameter is important since a smaller $\lambda$ will lead to high bias and low variance while a larger one leads to low bias and high variance of the proportion estimator. Based on this bias-variance trade-off, we take the optimal $\lambda$  to be the one that minimizes the MSE. 
Other researches in this direction include the use of change-point concepts for choosing $\lambda$ in the 
\cite{storey2002direct} estimator.
In this approach, first approximate the $p$-value plot by a piecewise linear function that has a single change-point. Select the $p$-value at this change-point location as the value of $\lambda$. Another approach is to choose $\lambda=\alpha$. \cite{hoang2021usage} noted that the default choice of $\lambda=0.5$ works well with randomized $p$-values since the sensitivity of the estimator in \eqref{schweder-def} with respect to $\lambda$ is least pronounced  for the case of randomized $p$-values. The default common choice of $\lambda=0.5$ is unstable, especially when dealing with dependent $p$-values. We plot Figure \ref{f:role_of_lambda} to illustrate how the estimator based on the UMP and RAND2 $p$-value is affected by different choices of $\lambda$. In this plot, we have used the same COVID-19 data and set $c=0.5$, $\theta_1=0.2963$, and $\theta_2=0.7566$.

From Figure \ref{f:role_of_lambda}, the estimate of $k_0$ based on the UMP $p$-value moves away from the number of true null hypotheses $k_0$ as the value of $\lambda$ increases. The estimate based on RAND2 $p$-value stays close to $k_0$ and only oscillates wildly around $k_0$ when $\lambda$ is close to $1$.

\section{Discussion}
\label{s:discuss}

In this research, we have considered the UMP and randomized $p$-value (RAND2) in the interval composite null hypothesis. Using large sample sizes, we have illustrated that the power functions for the UMP and RAND2 $p$-values are monotonically increasing in sample size. We have also found that it is possible for the power function of the UMP and the two-stage randomized $p$-value for a sample of size $n$ and that of $n+i$ for small $i\in \mathbb{N}$ to coincide on the entire parameter space. This problem occurs when dealing with relatively small samples for $\Delta$ too wide while the chosen parameter $\theta$ under the alternative is too far from the boundary. This problem does not occur when $\Delta$ is too narrow while the chosen parameter $\theta$ is too close to one of the boundaries of the alternative hypothesis (see Figure \ref{sample_power2}). This problem only occurs if the interval $\Delta$ gets smaller from one end while the other one is kept constant, for example, by holding $\theta_2$ constant and increasing $\theta_1$. 

The problem of nonmonotonicity of the power functions gets worse if the equivalence limit decreases from both ends. A similar observation in  \cite{qiu2010evaluation} is that when the equivalence limit is too narrow, the ROC curve of the TOST procedure is nonmonotonic for small sample sizes.
A complete characterization of this paradox will be considered in future research following the ideas in \cite{finner2001increasing} and \cite{finner1993behaviour}. Of course, in practical problems, the equivalence limits are determined before the data collection and remain fixed throughout the experiment. The adjustments made here are to illustrate the behavior of the $p$-values and their CDFs under different equivalence limits.

A plot of the CDFs for the UMP and RAND2 $p$-values under the null and alternative hypothesis illustrates that the UMP $p$-value is more conservative but less powerful compared to RAND2 $p$-value. The conservativeness of the UMP $p$-value increases while the one for RAND2 reduces with a further decrease in  $\Delta$. Furthermore, the power functions for the $p$-values are decreasing with $\Delta$. This decrease in power implies there is more benefit in using RAND2 as $\Delta$ reduces. A similar trend for the CDFs, which leads to the same conclusion, occurs when $\Delta$ is kept constant, and the chosen parameter under the null (alternative) is too far from (close to) the boundary of the null (alternative). 

Increasing both the parameters $\theta_1$ and $\theta_2$ by $\epsilon_1$ and $\theta$ by $\epsilon_2$ leads to an increase in power for both the $p$-values, a decrease in conservativity of the UMP $p$-value, and no change in the level of conservativity of RAND2 $p$-value. A similar trend occurs for a large equivalence limit, provided $\epsilon_2>\epsilon_1$. The power increases for a large equivalence limit since $\theta$ moves closer to the midpoint of $\Delta$, which is the parameter that gives the maximum power for both UMP and RAND2 $p$-values under this condition. We were also interested in finding the parameter value that maximizes the CDFs under the alternative hypothesis for the two $p$-values. We found that for large $\Delta$, the parameter value that maximizes the CDFs of both $p$-values occurs at the midpoint of $\Delta$. For small $\Delta$, however, this parameter value can be too close to the boundary of the alternative hypothesis for RAND2 $p$-value while the one for the UMP $p$-value is always at the midpoint.

Concerning the level of conservativity of the $p$-values to the sample size, we found that the CDF for the UMP $p$-value moves further away while the one for RAND2 $p$-value moves closer to the $UNI(0,1)$ line with an increase in the sample size. Therefore, the UMP $p$-value becomes more conservative while the level of conservativity for RAND2 $p$-value remains the same with an increase in the sample size. Furthermore, \cite{munk1996equivalence} and \cite{wellek2010testing} Sect. 1.2 (p. 5) argues that equivalence tests require much larger sample sizes to achieve a reasonable power compared to the one- or two-sided tests; unless $\Delta$ is chosen too wide that even “nonequivalent” hypotheses would be declared “equivalent.” Therefore, it would be better to consider  RAND2 $p$-value for multiple equivalence tests since they are less conservative even for large sample sizes.

A plot of the empirical CDFs of the $p$-values evaluates their performance when used with the estimator in \eqref{schweder-def}. The ECDF of RAND2 $p$-value, unlike the one for the UMP $p$-value, is above the $UNI(0,1)$ throughout. Furthermore, the slope between the points $(1,1)$ and $(\lambda, F_k^{RAND2})$ for RAND2 $p$-value is close to $\pi_0$ compared to the one for the UMP $p$-value. Therefore, RAND2 $p$-value outperforms the UMP $p$-value in estimating the proportion of true null hypotheses. To further justify this claim, we have given a real example and provided a simulation study, showing that RAND2 $p$-value outperforms the UMP $p$-value for all values of $\Delta$ by giving estimates that are closer on average to the true proportions.

The choice of the tuning parameter $\lambda$ for the estimator in \eqref{schweder-def} has also been of great concern in the recent literature. The sensitivity of this estimator to $\lambda$ is more pronounced for conservative $p$-values than for non-conservative ones. Since the UMP $p$-value is more conservative than RAND2 $p$-value, the choice of $\lambda$ is critical for obtaining estimates that are close to the actual number of true null hypotheses when using the UMP $p$-value.Based on the results from our simulation study, we recommend a small value of $\lambda$  when utilizing the UMP $p$-value. Assuming we are using a small $\alpha$, this choice is similar to the recommended choice of $\lambda=\alpha$ in the previous literature. When using RAND2 $p$-value, any choice of $\lambda$ which is not close to one is recommended. We recommend this choice since the estimate of $k_0$ based on RAND2 $p$-value oscillates wildly around the value of $k_0$ as $\lambda\longrightarrow 1$.

The recommendation in \cite{dickhaus2013randomized},\cite{habiger2011randomised}, and \cite{habiger2015multiple} that randomized $p$-values are nonsensical for a single hypothesis also applies to our RAND2 $p$-value and in that case the usage of the UMP $p$-value is advocated for. Furthermore, we caution the practitioner against using randomized $p$-values in bioequivalence studies. Some general extensions of this research include using randomized test procedures to achieve unbiased tests for Lehmann's alternative. Also, one could extend these procedures to consider multiple endpoints while accounting for the correlations among those endpoints. Finally, randomized $p$-values can be extended to stepwise regression since we use the $p$-values in these procedures several times, leading to multiple test problems.


\begin{appendices}

\section{Mathematical proofs}\label{secA1}
\setcounter{equation}{0}\renewcommand\theequation{A\arabic{equation}} 

\begin{proof}[Proof of Lemma~{\upshape\ref{max_p_value}}]
Recall that our (random) data is given by $\pmb{X}$, $U$ is a $ UNI(0,1)$-distributed random variable which is independent of our data, and $t\in [0,1]$ is an arbitrary significance level. Define 
$\phi(X,U;t)$ to be a decision function for a test procedure such that we reject the null when  $\phi(X,U;t)=1$ and otherwise fail to reject when $\phi(X,U;t)=0$. A $p$-value based on this decision function is 
\begin{equation}
P(X,U,)=\text{inf}\{t\in[0,1]:\phi(X,U;t)=1\}.
\label{p_val_def}
\end{equation}
Consider a test of $H:\theta\leq \theta_0$ versus $K:\theta>\theta_0$ where $\theta_0$ is a prespecified constant. The size of this test for an arbitrary $t\in[0,1]$ is 
\begin{equation}
E_{\theta_0}\{\phi_u(X,U;t)\}=\mathbb{P}_{\theta_0}(T(\pmb{X})> C_n)+\gamma_n\mathbb{P}_{\theta_0}(T(\pmb{X})=C_n)=t,
\end{equation}
where $C_n=F^{-1}_{\theta_0}(1-t)$ and $\gamma_n=\{\mathbb{P}_{\theta_0}(T(\pmb{X})\leq C_n)-(1-t)\}\{\mathbb{P}_{\theta_0}(T(\pmb{X})=C_n)\}^{-1}$ are the critical and randomization constants, respectively as given in Definition \ref{def_ump_p_value}. The $p$-value for this test based on the definition in Equation \eqref{p_val_def} is
\[P_u(X,U)=\text{inf}\{t\in[0,1]:t\geq \mathbb{P}_{\theta_0}(T(\pmb{x})> c_n)+u\mathbb{P}_{\theta_0}(T(\pmb{x})=c_n)\},\]
\[=\mathbb{P}_{\theta_0}(T(\pmb{X})> C_n)+U\mathbb{P}_{\theta_0}(T(\pmb{X})=C_n).\]
Similarly, consider a test of the form $H:\theta\geq \theta_0$ versus $K:\theta<\theta_0$ where again $\theta_0$ is a prespecified constant. The size of this test for an arbitrary $t\in[0,1]$ is 
\begin{equation}
E_{\theta_0}\{\phi_l(X,U;t)\}=\mathbb{P}_{\theta_0}(T(\pmb{X})\leq D_n-1)+\delta_n\mathbb{P}_{\theta_0}(T(\pmb{X})=D_n)=t,
\end{equation}
where again $D_n=F^{-1}_{\theta_0}(t)$ and $\delta_n=\{t-\mathbb{P}_{\theta_0}(T(\pmb{X})\leq D_n-1)\}\{\mathbb{P}_{\theta_0}(T(\pmb{X})=D_n)\}^{-1}$ are the critical and randomization constants, respectively as given in Definition \ref{def_ump_p_value}.
The $p$-value for this test based on the definition in Equation \eqref{p_val_def} is
\[P_l(X,U)=\text{inf}\{t\in[0,1]: t\geq \mathbb{P}_{\theta_0}(T(\pmb{x})\leq d_n-1)+u\mathbb{P}_{\theta_0}(T(\pmb{x})=d_n)\},\]
\[=\mathbb{P}_{\theta_0}(T(\pmb{X})\leq D_n-1)+U\mathbb{P}_{\theta_0}(T(\pmb{X})=D_n).\]
Assume now that the hypothesis is as given in Equation \eqref{testproblem}, then the overall test statistic for this problem is
\[\phi_{m}(X,U;t)=\text{min}\{\phi_l(X,U;t),\phi_u(X,U;t)\},\]
\[=\phi_l(X,U;t)\bigcap \phi_u(X,U;t).\]
The overall $p$-value is 
\begin{equation}   
P_{m}(X,U)=\text{max}\{P_{l}(X,U),P_{u}(X,U)\},
\label{p_overall}
\end{equation} since
\[\{t:\phi_{m}(X,U;t)=1\}=
\{t:\text{min}[\phi_{l}(X,U;t), \phi_{u}(X,U;t)]=1\},\]
\[=
\{[t:\phi_{l}(X,U;t)=1]\bigcap [t:\phi_{u}(X,U;t)=1]\},\]
\[=\{t:t \geq \mathbb{P}_{\theta_0}(T(\pmb{x})\leq d_n-1)+u\mathbb{P}_{\theta_0}(T(\pmb{x})=d_n)\} \bigcap\{t:t \geq \mathbb{P}_{\theta_0}(T(\pmb{x})> c_n)+u\mathbb{P}_{\theta_0}(T(\pmb{x})=c_n)\},\]
\[=\{t:t\geq P_{l}(X,U)\}\bigcap\{t:t\geq P_{u}(X,U)\},\]
\[=\{t:t\geq \text{max}[ P_{l}(X,U),P_{u}(X,U)]\},\]
which gives the overall $p$-value in \eqref{p_overall} using the definition in \eqref{p_val_def}. We can express this further as
\[\{t:\phi_{m}(X,U;t)=1\}=\{t:t \geq \mathbb{P}_{\theta_0}(c_n<T(\pmb{x})<d_n)+u\mathbb{P}_{\theta_0}(T(\pmb{x})=c_n)+u\mathbb{P}_{\theta_0}(T(\pmb{x})=d_2)\},\]
which gives the $p$-value 
\[P_{m}(X,U)=\mathbb{P}_{\theta_0}(C_n<T(\pmb{X})<D_n)+U\mathbb{P}_{\theta_0}(T(\pmb{X})=C_n)+U\mathbb{P}_{\theta_0}(T(\pmb{X})=D_2)\},\] again based on the definition of a $p$-value in Equation  \eqref{p_val_def}. However, this is equivalent to the UMP $p$-value in Definition \ref{def_ump_p_value} since $\theta_0$ is the LFC parameter, which can be $\theta_1$ or $\theta_2.$
\end{proof}

\begin{proof}[Proof of Theorem~{\upshape\ref{theorem:4.1}}]

To verify that the CDFs of the UMP and the two-stage randomized $p$-values are point-wise monotonically increasing with an increase in the sample size for any parameter value $\theta$ belonging to the alternative hypothesis, it suffices to prove that these CDFs for a sample of size $n+1$ are greater than those for size $n$. 
Recall from Equation \eqref{eq:rand2_p_value_cdf} that
\begin{eqnarray*}
\mathbb{P}_{\theta}\{P^{rand2}(\pmb{X},U,\Tilde{U},c)\leq t\} &= &t\mathbb{P}_{\theta}\{P^{UMP}(\pmb{X},U)> c\}+\mathbb{P}_{\theta}\{P_T^{UMP}(\pmb{X},U)\leq tc\},\\
&= &t-t\mathbb{P}_{\theta}\{P^{UMP}(\pmb{X},U)\leq c\}+\mathbb{P}_{\theta}\{P^{UMP}(\pmb{X},U)\leq tc\}.    
\end{eqnarray*}
For an arbitrary, but fixed $c\in[0,1]$, further recall that $C_n=F_{Bin(n,\theta_1)}^{-1}(1-c)$ and $D_n=F_{Bin(n,\theta_2)}^{-1}(c)$ denotes the quantile of a binomial random variable with parameters $n$ and $\theta_i$ for $i=1,2$.
Again recall that the randomization constants are given by \[\gamma_n=\dfrac{\mathbb{P}_{\theta_1}(T(\pmb{X})\leq C_n)-(1-c)}{\mathbb{P}_{\theta_1}(T(\pmb{X})=C_n)}\ \text{and}\ \delta_n=\dfrac{c-\mathbb{P}_{\theta_2}(T(\pmb{X})\leq D_n-1)}{\mathbb{P}_{\theta_2}(T(\pmb{X})=D_n)}.\]
Define 
\[
\beta_n \equiv \beta_n(c, \theta, \theta_1,\theta_2) = 
\mathbb{P}_{\theta}\{ P^{UMP}(\pmb{X})\leq c\},\]
\begin{equation}
=\mathbb{P}_{\theta}(C_n<T(\pmb{X})<D_n)+\gamma_n\mathbb{P}_{\theta}(T(\pmb{X})=C_n)+\delta_n\mathbb{P}_{\theta}(T(\pmb{X})=D_n).
\label{eq:a0}
\end{equation}
Since $X$ is a random variable that follows a binomial distribution with parameters $n$ and $\theta$, the above power function can be expressed as
\begin{equation}
\beta_n=(1+\varrho)^{-n}
\bigg\{
\sum_{x={c_n}+1}^{n} {n\choose x}\varrho^x
+\gamma_n{n\choose c_n}\varrho^{c_n}-\sum_{x=d_n}^{n} {n\choose x}\varrho^x+\delta_n{n\choose d_n}\varrho^{d_n}
\bigg\},
\label{eq:a1}
\end{equation}
where $\varrho=\bigg(\dfrac{\theta}{1-\theta}\bigg)$ and it is such that $\varrho\in(0,\infty)$. For a sample of size $n+1$, Equation \eqref{eq:a1} becomes
\begin{equation}
\beta_{n+1}=(1+\varrho)^{-n-1}
\bigg\{
\sum_{x={c_{n+1}}+1}^{n+1} {n+1\choose x}\varrho^x
+\gamma_{n+1}{n+1\choose c_{n+1}}\varrho^{c_{n+1}}-\sum_{x=d_{n+1}}^{n+1} {n+1\choose x}\varrho^x+\delta_{n+1}{n+1\choose d_{n+1}}\varrho^{d_{n+1}}
\bigg\}.
\label{eq:a2}
\end{equation}
Since $\gamma_n,\gamma_{n+1},\delta_n,\delta_{n+1}\in (0,1)$,
to verify that $\beta_{n+1}>\beta_n$, we compare the coefficients of $\varrho^x$ in Equations \eqref{eq:a1} and \eqref{eq:a2}. To do this for the coefficients of $\varrho^x$ in the first terms in Equations \eqref{eq:a1} and \eqref{eq:a2}, we have
\begin{equation}
(1+\varrho)\bigg[\sum_{x=c_n+1}^{n} {n\choose x}\varrho^x\bigg]<\bigg[\sum_{x=c_{n+1}+1}^{n+1} {n+1\choose x}\varrho^x\bigg],
\label{eqn03}
\end{equation}
provided $c_{n}=c_{n+1}.$ The proof for the other case when $c_{n}+1=c_{n+1}$ can be shown similarly. 
Next, comparing the other coefficients of $\varrho^x$ in Equations \eqref{eq:a1} and \eqref{eq:a2}, we have
\begin{equation}
(1+\varrho)\bigg[\sum_{x=d_n}^{n} {n\choose x}\varrho^x\bigg]>\bigg[\sum_{x=d_{n+1}}^{n+1} {n+1\choose x}\varrho^x\bigg],
\label{eqn03}
\end{equation}
provided $d_{n}+1=d_{n+1}.$ Again proving the other case when $d_{n}=d_{n+1}$ will follow similar steps. With this, it is evident that $\beta_{n+1}>\beta_n$. The proof for $\mathbb{P}_{\theta}\{P_T^{rand}(\pmb{X},U)\leq tc\}$, which yields the same result, can be carried out similarly. With this, the CDF for the two-stage randomized $p$-value $RAND2$ is, under the stated conditions, monotonically increasing with an increase in $n$, which we needed to prove. Repeating the above calculations for $\beta_n$ with $t\in[0,1]$ in place of $c$ provides the proof that the CDF of the UMP $p$-value is monotonically increasing with an increase in the sample size (under the stated conditions).
\end{proof}
\end{appendices}

\bibliography{Main.bbl.bib}


\end{document}